\documentclass{iopart}
\usepackage{amssymb}
\usepackage{graphicx}

\begin{document}

\title{Thermodynamics and Cardy-like formula for nonminimally dressed, charged Lifshitz black holes in New Massive Gravity}

\author{Mois\'es Bravo-Gaete}
\ead{mbravo-at-ucm.cl} \address{Facultad de Ciencias
B\'asicas, Universidad Cat\'olica del Maule, Casilla 617, Talca,
Chile.}

\author{Mar\'ia Montserrat Ju\'arez-Aubry}
\ead{mjuarez-at-fis.cinvestav.mx} \address{Arkansas State University Campus Quer\'etaro,
 Carretera estatal $\#100$ km. 17.5, Municipio Col\'on, 76270, {Quer\'etaro}, M\'exico}

\begin{abstract}
{In three dimensions, we consider the Einstein-Maxwell Lagrangian dressed by a nonminimally coupled scalar field in New Massive Gravity. For this theory, we provide two families of electrically charged Lifshitz black holes where their metric functions depend only on an integration constant.} We calculate their masses using the quasilocal approach, as well as their entropy and electric charge. {These charged configurations are interpreted as extremal in the sense that the mass vanishes identically while the entropy and electric charge are non zero thermodynamic quantities.} Using these examples, {we corroborate that the semiclassical entropy can be recovered through a charged} Cardy-like formula, involving the corresponding magnetically charged solitons {obtained by a double Wick rotation.} Finally, the first law of thermodynamics, as well as the Smarr formula are also verified.
\end{abstract}

\maketitle
\newpage

\section{Introduction}

In recent years, the prospect of extending the idea of the AdS/CFT correspondence \cite{Maldacena:1997re} to non-relativistic physics has been gaining momentum with the purpose of better understanding other systems such as condensed matter physics. One particularly interesting scenario is that of systems that have a dynamical scaling near fixed points. These systems are characterized by having a different scaling between the time $t$ and the space $\vec{x}$. In this instance, {under the gravity side,} Kachru et al proposed the Lifshitz {spacetime} \cite{Kachru:2008yh} which in three dimensions takes the form
\begin{equation}\label{eq:Lifshitz}
ds^2=-\frac{r^{2z}}{l^{2z}}dt^2+\frac{l^2}{r^2}dr^2+\frac{r^2}{l^2} d\varphi^{2},
\end{equation}
where $z$ is the  dynamical exponent responsible of the anisotropy of the scaling
\begin{equation*}
t\mapsto\tilde{\lambda}^z t, \quad r\mapsto\tilde{\lambda}^{-1}r, \quad \varphi\mapsto\tilde{\lambda}\varphi,
\end{equation*}
with {$0<t<+\infty,\, r \geq 0$ and} $0<\varphi<2\pi l$. Note that $z=1$ in eq. (\ref{eq:Lifshitz}) corresponds to the AdS spacetime.  Another aspect that is vital to the idea of a Gauge/Gravity correspondence is the finite temperature effects, through the introduction of black holes whose asymptotic behaviour reproduce the spacetime (\ref{eq:Lifshitz}), denominated as Lifshitz black holes. Unfortunately, we have come to learn that General Relativity with a cosmological constant is not enough to support the Lifshitz spacetime with $z \neq 1$ \cite{Kachru:2008yh} (and consequently, cannot support Lifshitz black holes either), this is why many approaches to find these configurations consider one of the following mechanisms: adding matter fields \cite{Pang:2009pd,Mann:2009yx,Alvarez:2014pra,BravoGaete:2019rci,Maeda:2011jj,Bravo-Gaete:2013dca,Tarrio:2011de,Taylor:2008tg,Danielsson:2009gi,Azeyanagi:2009pr,Herrera:2017ztd,Zangeneh:2015uwa}, enriching the gravitational Lagrangian with corrections in the curvature \cite{Cai:2009ac,AyonBeato:2009nh,AyonBeato:2010tm,Dehghani:2010kd,Matulich:2011ct,Lu:2012xu,Giacomini:2012hg} or both simultaneously \cite{Correa:2014ika,Ayon-Beato:2015jga,Bravo-Gaete:2015xea,Fan:2014ala,Ayon-Beato:2019kmz}.

In particular, {in 2+1 dimensions, there exists a ghost-free, parity even gravity theory} called New Massive Gravity (NMG) \cite{Bergshoeff:2009hq}, yielding a Lifshitz black hole solution characterized by a dynamical exponent $z=3$ \cite{AyonBeato:2009nh}.  Remarkably, it has been shown that the spectrum of values for the dynamical exponent $z$ becomes enriched by the introduction of a scalar field nonminimally coupled as a matter field to NMG \cite{Correa:2014ika,Ayon-Beato:2015jga}. Nevertheless, the charged case for this theory {still remains an open problem.}

Independently, in the context of the relevance of thermodynamical properties of Lifshitz black holes, one can point out that studying such configurations in three dimensions can serve as a good laboratory in order to study relevant aspects of the Gauge/Gravity duality (see for example \cite{Erices:2019onl}). In particular, the semiclassical entropy of these configurations can be recovered through a Cardy-like expression \cite{Gonzalez:2011nz} characterized by the mass of the black hole and the corresponding solitonic configuration by operating a double Wick rotation. Following the same spirit, in \cite{Bravo-Gaete:2015iwa} a charged Cardy-like formula was proposed. In the present work, we test the viability of this expression considering the NMG and a matter contribution given by a nonminimally coupled scalar field and a linear Maxwell field in three dimensions. We hope this study will contribute to enrich the list of charged Lifshitz black holes and that it will highlight the importance of the role played by the now charged soliton.

The rest of the paper will be organized as follows. In the next section, we present the action in three dimensions constituted by NMG-Maxwell theory together with a nonminimally coupled scalar field. We also present the corresponding equations of motion as well as a specific ansatz for charged Lifshitz black hole solutions. By using these ansatze, in section \ref{section:eom} we build two new charged asymptotically Lifshitz black holes. In section \ref{thermo}, we find a general expression for the Wald entropy and the mass of the electrically charged Lifshitz black holes together with the mass of their magnetically charged solitonic counterparts. Lastly, for our newly found solutions, we present these thermodynamical quantities checking the validity of a charged Cardy-like formula which reproduces the Wald entropy. In all these cases we verify the fulfillment of the charged version of the first law of black hole thermodynamics as well as  the anisotropic version
of the Smarr formula.  Finally, in section \ref{section:conclusions} we present our discussion and conclusions.

\section{Action and field equations}\label{actioneom}

{With all the above, and motivated with this rich list of exact Lifshitz black holes, in this work we look forward to finding more examples of asymptotically Lifshitz black holes, but now electrically charged. As a result,
we will} consider the 2+1 NMG theory \cite{Bergshoeff:2009hq}
%
together with a nonminimally coupled
scalar field $\Phi$  and the addition of {a linear Maxwell field}, namely
\begin{eqnarray}
\label{eq:Squad}
S[g_{\mu\nu},\Phi,A_{\mu}]&=& \int{d}^3 x\sqrt{-g}\, \,  \mathcal{L}
=\int{d}^3 x\sqrt{-g}\left(\mathcal{L}_{\mathrm{NMG}}+\mathcal{L}_{\mathrm{M}}\right),
\end{eqnarray}
with
\begin{eqnarray*}
&&\mathcal{L}_{\mathrm{NMG}}=\frac{1}{2\kappa}\left[ R-2\lambda - \frac{1}{m^2} \left( R_{\mu\nu}R^{\mu\nu}-\frac{3}{8}R^2\right)  \right],\nonumber \\
&&\mathcal{L}_{\mathrm{M}}=-\frac{1}{2}\nabla_{\mu}\Phi\nabla^{\mu}\Phi-\frac{\xi}{2}R\Phi^2-U(\Phi)-\frac{1}{4} F_{\mu\nu}F^{\mu\nu},
\end{eqnarray*}
where $\Phi$ is the scalar field,  $U(\Phi)$ is the self interaction potential, $\xi$ represents the non-minimal coupling parameter and $F_{\mu\nu}\equiv 2\partial_{[\mu}A_{\nu]}$ is the strength of the Maxwell field $A_{\mu}$.

The field equations obtained by varying the action with respect to the
metric $g_{\mu\nu}$, the scalar field $\Phi$ and $A_{\mu}$ read
\begin{eqnarray}
E_{\mu\nu}\equiv G_{\mu\nu}+\lambda g_{\mu\nu}-\frac{1}{2m^2}K_{\mu\nu}-\kappa T_{\mu\nu}=0, \label{eqmotion1}\\
\Box\Phi - \xi R\Phi = \frac{dU(\Phi)}{d\Phi},\label{eqmotion2}\\
\nabla_{\mu}F^{\mu\nu}=0, \label{eq:Maxwelleqns}
\end{eqnarray}
where we have defined
\begin{eqnarray}
K_{\mu\nu}={}&2\Box R_{\mu\nu} -\frac{1}{2}
\left( g_{\mu\nu}\Box+\nabla_{\mu}\nabla_{\nu}-9R_{\mu\nu} \right)R
\nonumber\\
&-8R_{\mu\alpha}R^{\alpha}_{~\nu}
+g_{\mu\nu}\left( 3R^{\alpha\beta}R_{\alpha\beta}-\frac{13}{8}R^2 \right),
\end{eqnarray}
and the energy-momentum tensor is given  by
\begin{eqnarray}
T_{\mu\nu}=&&\nabla_{\mu}\Phi\nabla_{\nu}\Phi
-g_{\mu\nu}\left(\frac{1}{2}\nabla_{\sigma}\Phi\nabla^{\sigma}\Phi+U(\Phi)\right)\nonumber\\
&&+\xi\left( g_{\mu\nu}\Box - \nabla_{\mu}\nabla_{\nu}+G_{\mu\nu}
\right)\Phi^2\nonumber\\
&&+F_{\mu\sigma}F^{\sigma}_{\nu}-\frac{1}{4}g_{\mu\nu}F_{\alpha\beta}F^{\alpha\beta}.
\end{eqnarray}
 As was shown in \cite{AyonBeato:2009nh}, in absence of matter sources, the equations of motion (\ref{eqmotion1})  support the Lifshitz spacetime (\ref{eq:Lifshitz}) for a generic value of the dynamical exponent $z$ when
\begin{eqnarray}
m^2&=&-\frac{z^2-3z+1}{2l^2}, \label{eq:m2}\\
\lambda&=&-\frac{z^2+z+1}{2l^2}, \label{eq:lambda}
\end{eqnarray}
where values of $z$ giving a vanishing value for $m^2$ and $\lambda$ are forbidden.
Moreover, since we are interested in looking for charged black hole solutions that asymptote
the Lifshitz spacetime (\ref{eq:Lifshitz}), we not only impose (\ref{eq:m2}) and (\ref{eq:lambda}), but we also opt for the
following ansatz
\begin{eqnarray}
\label{lifbh} && ds^2 = - \frac{r^{2 z}}{\l^{2 z}} f(r) dt^2 +
\frac{\l^2}{r^2} \frac{dr^2}{f(r)} +
  \frac{r^2}{\l^2} d\varphi,\\
&& \Phi=\Phi(r),
\end{eqnarray}
with $\lim_{r
\rightarrow +\infty} f(r)=1$ and we supplement our suppositions by considering an electrical ansatz for the electromagnetic field $A_{\mu}=A_{t}(r)\delta^t_{\mu}$. Under these assumptions,
the Maxwell's equations (\ref{eq:Maxwelleqns}) together with the line element (\ref{lifbh}) give
$$\left[\left(\frac{r}{l}\right)^{2-z} \,A'_{t}\right]'=0,$$
where $(')$ denotes the derivative with respect the radial coordinate $r$. This expression is integrated straightforwardly giving
\begin{equation}\label{eq:vecpot}
A_{t}(r)=\frac{q\,r_h\,r^{z-1}}{(z-1)l^{z-2}}.
\end{equation}
The integration constant in our case is null while the electric field strength reads
\begin{equation}\label{eq:strength}
F_{rt}=A'_{t}=q r_h \left(\frac{r}{l}\right)^{z-2},
\end{equation}
where ${q}$ is a real constant and $r_h$  is an integration constant related to the location of the event horizon. While the power of $r$ is a direct consequence of fulfilling Maxwell's equations, this particular form of (\ref{eq:vecpot}) (or (\ref{eq:strength})) is convenient for simplifying the computations. Equations (\ref{lifbh}) to (\ref{eq:strength}) will constitute the starting point of our task of looking for asymptotically Lifshitz black holes for the action (\ref{eq:Squad}). \\

\section{Two classes of Charged Lifshitz Black Holes} \label{section:eom}

Let us begin by considering how the equations of motion (\ref{eqmotion1}) are affected by the electric source. The fact that the mixed temporal and radial components of the Maxwell energy-momentum tensor coincide implies that the analysis of the combination $E^t_{\, t}-E^r_{\, r}=0$ is a good starting point for our study. Indeed, this combination yields to a fourth order Cauchy-Euler differential equation in $f(r)$, plus a non-linear contribution of $\Phi(r)$. More explicitly, this combination is proportional to
\begin{eqnarray*}
&&r^4 f^{(4)} +2(z+4)r^3f''' -(z^2-17z-8)r^2 f''\nonumber\\
&&-2(z+2)r(z^2-5z+2)f'-2(z-1)(z^2-3z+1)f
\nonumber \\
&&-8\Phi \Phi''\kappa l^2 r^2 \xi m^2 - 4\kappa l^2 m^2 r^2 (2\xi-1)\big(\Phi'\big)^2+ 8\kappa l^2 m^2 r\xi (z-1)\Phi \Phi '\nonumber\\
&&+4\kappa l^2 m^2\xi (z-1)\Phi^2- 4 m^2 l^2(z-1) = 0,
\end{eqnarray*}
where, as before, $(')$ denotes the derivative with respect the radial coordinate $r$. As a result, we choose the following ansatze for the metric function and the scalar field
\begin{eqnarray}
f(r)=1-\left(\frac{r_h}{r}\right)^{\chi},\qquad
\Phi(r)=\sqrt{\Phi_{0}} \left(\frac{r_h}{r}\right)^{\gamma},\label{gensolLif}
\end{eqnarray}
where $\chi$ and $\gamma$ are nonnegative constants and, as before, $r_h$ denotes the location
of the horizon while $\Phi_0$ is a positive constant.
As a result, the combination ${E^t_{\, t}-E^r_{\, r}}=0$, with $m^2$ given previously in (\ref{eq:m2}), yields
\begin{eqnarray}
&&2 \kappa \Phi_0  (z^2-3 z+1)   P_2(\gamma,\xi,z)   \left(\frac{r_h}{r}\right)^{2 \gamma}- P_4(\chi,z) \left(\frac{r_h}{r}\right)^\chi=0,
\end{eqnarray}
with
\begin{eqnarray}
P_2(\gamma,\xi,z)&=& (4 \xi -1) \gamma^2+2 \gamma \xi  z+\xi  (1-z), \nonumber\\
P_4(\chi,z)&=&\chi^4-(2 z+2) \chi^3-(z^2-11 z+5) \chi^2\nonumber \\
&+&(z+1) (2 z^2-9 z+6) \chi-2 (z-1) (z^2-3 z+1). \nonumber
\end{eqnarray}
{When looking for Lifshitz solutions ($z\neq 1$), the simultaneous vanishing of both $P_2(\gamma,\xi,z)$ and $P_4(\chi,z)$ will yield to inconsistencies when substituted into the other equations of motion (\ref{eqmotion1})-(\ref{eq:Maxwelleqns}) except for the known cases without charge cited in \cite{Ayon-Beato:2015jga} and the vacuum solution of $z=3$ \cite{AyonBeato:2009nh} (proved to be partially unique in \cite{Ayon-Beato:2014wla}).} As a result, we must consider the possibility of having $\gamma=\chi/2$, which fixes the value of $\Phi_0$ to
\begin{eqnarray}\label{eq:phi0}
\Phi_0&=&\frac{P_4(\chi,z)}{2\kappa (z^2-3 z+1)P_2(\chi/2,\xi,z)}.
\end{eqnarray}
As such let us now capitalize on the fact that we have a new more restricted ansatz
\begin{eqnarray}
f(r)=1-\left(\frac{r_h}{r}\right)^{\chi},\qquad
\Phi(r)=\sqrt{\Phi_{0}} \left(\frac{r_h}{r}\right)^{\chi/2}.\label{gensolLif_chi}
\end{eqnarray}

In addition to the equations $E^t_{\, t}-E^{r}_{\, r}=0$, where the metric function and the scalar field are {given} in (\ref{gensolLif_chi}) {and} the constant $\Phi_0$ is given in (\ref{eq:phi0}), the combination of $E^t_{\, t}-E^{\varphi}_{\, \varphi}=0$ yields
\begin{eqnarray}\label{eq:ett-exx}
&& \frac{\Upsilon_1}{z^2-3 z+1}   \left(\frac{r_h}{r}\right)^{\chi}
- \frac{\Upsilon_2}{z^2-3 z+1}  \left(\frac{r_h}{r}\right)^{2\chi} \nonumber\\
&&+4 l^2 q^2 \kappa \left(\frac{r_h}{r}\right)^{2}=0,
\end{eqnarray}
where
\begin{eqnarray*}
\Upsilon_1&=&2 (\chi-z-1) [2 \chi^3-(5 z-1) \chi^2+(z+6) (z-1) \chi\nonumber \\
&&-2 (z-1) (z^2-3 z+1) (\kappa \Phi_0 \xi-1)],\\
\Upsilon_2&=&(2 \chi-z-1) (\chi-2 z+2) [3 \chi^2-(z+6) \chi \nonumber \\
&&+2 (z^2-3 z+1) (\kappa \Phi_0 \xi-1)].
\end{eqnarray*}
{Furthermore,  with an election of the potential $U(\Phi)=\sigma_1 \Phi^4+\sigma_2 \Phi^2$ (with $\sigma_i$ being coupling constants) the remaining independent Einstein equation $E_{\varphi}^{\varphi}=0$ reads
\begin{eqnarray}\label{Eii}
\frac{\Theta_1}{(z^2-3z+1)l^2}\left(\frac{r_h}{r}\right)^{\chi}
+\frac{\Theta_2}{(z^2-3z+1)l^2} \left(\frac{r_h}{r}\right)^{2 \chi}\nonumber\\
-\frac{l^2 q^2 \kappa}{2} \left(\frac{r_h}{r}\right)^{2}+\lambda+\frac{z^2+z+1}{2l^2}=0,
\end{eqnarray}
where
\begin{eqnarray*}
\Theta_1&=&{\frac {7}{8}}\,{\chi}^{4}-\frac{11}{4}\,z{\chi}^{3}-\left[ \frac{1}{8}\,{\Phi_0}\,\kappa\, \left( {z}^{2}-3\,z+1 \right)  \left( 20\,\xi-1
 \right) +4\,z\right.\\
 &+&\left.{\frac {15}{8}}\,{z}^{2}-4 \right] {\chi}^{2}
 -\frac{1}{2}\,z
 \left[ 5\,\kappa\,\xi\,{\Phi_0}\, \left( {z}^{2}-3\,z+1 \right) -{z
}^{2}+12\,z-12 \right] \chi\\
&+&\kappa\,{l}^{2}{\sigma_1} \left( {z}^{2}-3\,z+1 \right)
\Phi_0^{2}
+\kappa\,\xi\,{z}^{2} \left( {z}^{2}-3\,z+1
 \right) {\Phi_0}\\
 &-&\frac{1}{2}\, \left( {z}^{2}-3\,z+1 \right)  \left( {z}^{2
  }-z-1 \right),\\ \\
  \Theta_2&=& -\frac{1}{2}\,{\chi}^{4}+2\,z{\chi}^{3}+ \left( -\frac{5}{2}
\,{z}^{2}+\frac{3}{2}-\frac{1}{2}\,z \right) {\chi}^{2}+\frac{1}{2}\,z \left( z+3 \right)\left( 2\,z-3 \right) \chi\\
&-&\frac{1}{8}\,{\Phi_0}\,\kappa\, \left( {z}^{2}-3\,z+1 \right)  \left[ 8\,
\xi\, \left( {z}^{2}-z\chi+{\chi}^{2} \right) -{\chi}^{2}-8\,{\sigma_2}\,{l}^{2} \right]\\
&-&\left( z+1 \right)  \left( {z}^{2}-3\,z+1
\right),
\end{eqnarray*}
with the cosmological constant $\lambda$ fixed as (\ref{eq:lambda}). It is clear, from eqs. (\ref{eq:ett-exx})-(\ref{Eii})  } that the only way to obtain charged solutions is to consider either $\chi=1$ or $\chi=2$, which is consistent with the general expression of the mass of the black hole (\ref{massgensol}).

\subsection{Solution with arbitrary dynamical exponent z}

The first case that we consider is for $\chi=1$. In this particular scenario, the non-minimal coupling is fixed to the value
\begin{eqnarray}\label{xisol1}
\xi = \frac{2z^3-7z^2+8z-5}{16(z-2)(z^2-z+1)},
\end{eqnarray}
and we find a solution given by
{\begin{eqnarray}
ds^2&=&-\frac{r^{2z}}{l^{2z}} \left( 1- \frac{r_h}{r} \right) dt^2
+\frac{l^2}{r^2} \left( 1- \frac{r_h}{r} \right)^{-1} dr^2 \nonumber\\
&&+\frac{r^2}{l^2} d\varphi^{2}, \label{sol:chi1} \\
\Phi (r) &=& \sqrt{ \frac{ 8(z^2-z+1)(z-2)}{(z-1)(z^2-3z+1)\kappa}} \sqrt{\frac{r_h}{r}}, \\
F_{rt}&=&\sqrt{-\frac{z(2z-3)}{2\kappa(z^2-3z+1)l^4}}\, \, r_h\left(\frac{r}{l}\right)^{z-2},
\end{eqnarray}}
and this solution is supported by the self-interaction potential
\begin{eqnarray}
U( \Phi) &=&- {\frac {\kappa  \left( {z}^
{2}-3 z+1 \right)
   \left( z-1 \right) ^{2}
\tilde{P}_4(z)
}{512\, {l}^{2} \left( {z}^{2}-z+
1 \right) ^{2} \left( z-2 \right) ^{2}}}
{\Phi}^{4}
\nonumber \\
&&+ {\frac { \left( z-1 \right)  \left( {z}^{2}-3 z+1 \right)  \left( 2 {z}^{2}-z+1
 \right) }{ 16\,  \left( {z}^{2}-z+1 \right)  \left( z-2 \right) {
l}^{2}}} {\Phi}^{2},\label{eq:potsol1}
\end{eqnarray}
with
\begin{eqnarray}
\tilde{P}_4(z)&=&4 {z}^{4}-20 {z}^{3}+33 {z}^{2}-28 z+13,
\end{eqnarray}
while the remaining constants are given by eqs.(\ref{eq:m2}) and (\ref{eq:lambda}).
In order to ensure the realness of the solution, we must consider the range of $z$ in accordance to TABLE \ref{tabla1}.

\begin{table}[h!]
\begin{center}
\begin{tabular}{|c|c|}
\hline
$\kappa$ & Range of $z$ \\
\hline \hline
~$\kappa>0$~ & ~$z\in(0,\frac{3-\sqrt{5}}{2}\approx 0.382) \cup (\frac{3}{2},2)$~ \\  [1ex]
\hline \hline
~$\kappa<0$~ & ~$z\in ( \frac{3-\sqrt{5}}{2}\approx 0.382,1) $~ \\  [1ex]
\hline
\end{tabular}
\end{center}
\caption{\label{tabla1}Range of possibilities for the dynamical
exponent $z$ allowing black holes.}
\end{table}%

\begin{figure}[h!]
  \centering
    \includegraphics[scale=.42]{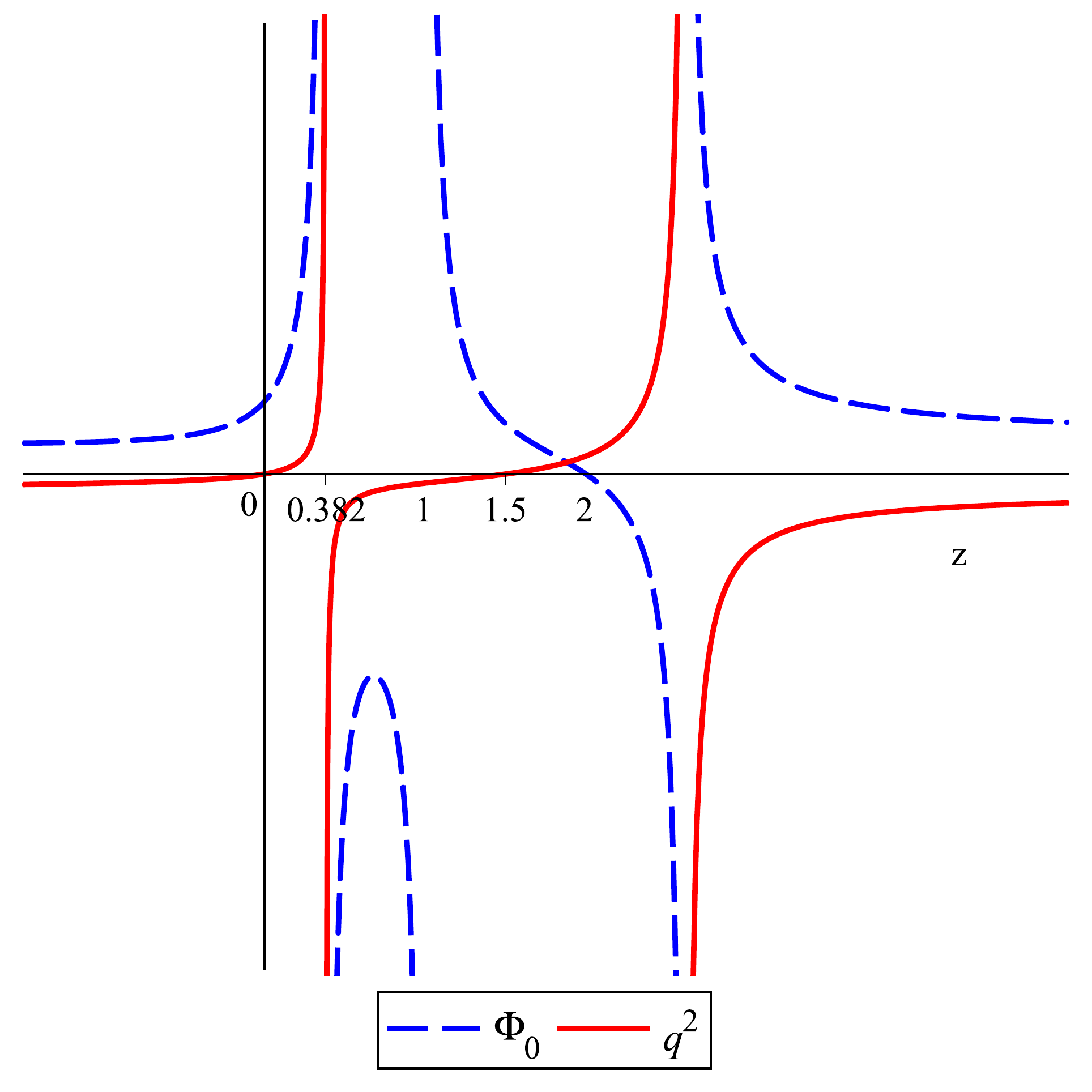}
      \caption{Representation of $q^2$ and $\Phi_0$ as functions of $z$ for the $\chi=1$ configuration, assuming $\kappa>0$.}
  \label{fig:q2vsphi0_chi1}
\end{figure}

Moreover, FIG. \ref{fig:q2vsphi0_chi1}  is provided as a visual aid on this matter. A quick inspection of said figure shows that the intervals $z\in (0,\frac{3-\sqrt{5}}{2})$ and $z\in(3/2,2)$ are the only ones for which
$${\Phi_0=\frac{ 8(z^2-z+1)(z-2)}{(z-1)(z^2-3z+1)\kappa}} \quad \mbox{and}
\quad q^2=-\frac{z(2z-3)}{2\kappa(z^2-3z+1)l^4},$$
are simultaneously positive for $\kappa>0$. Notice that taking $\kappa<0$ (equivalent to a reflection with respect to the horizontal axis) will provide us with the interval $z\in (\frac{3-\sqrt{5}}{2},1)$ in which $\Phi_0$ and $q^2$ are simultaneously positive.

\subsection{Solution with a fixed value of the dynamical exponent z=2}
\begin{figure}[h!]
  \centering
    \includegraphics[scale=.42]{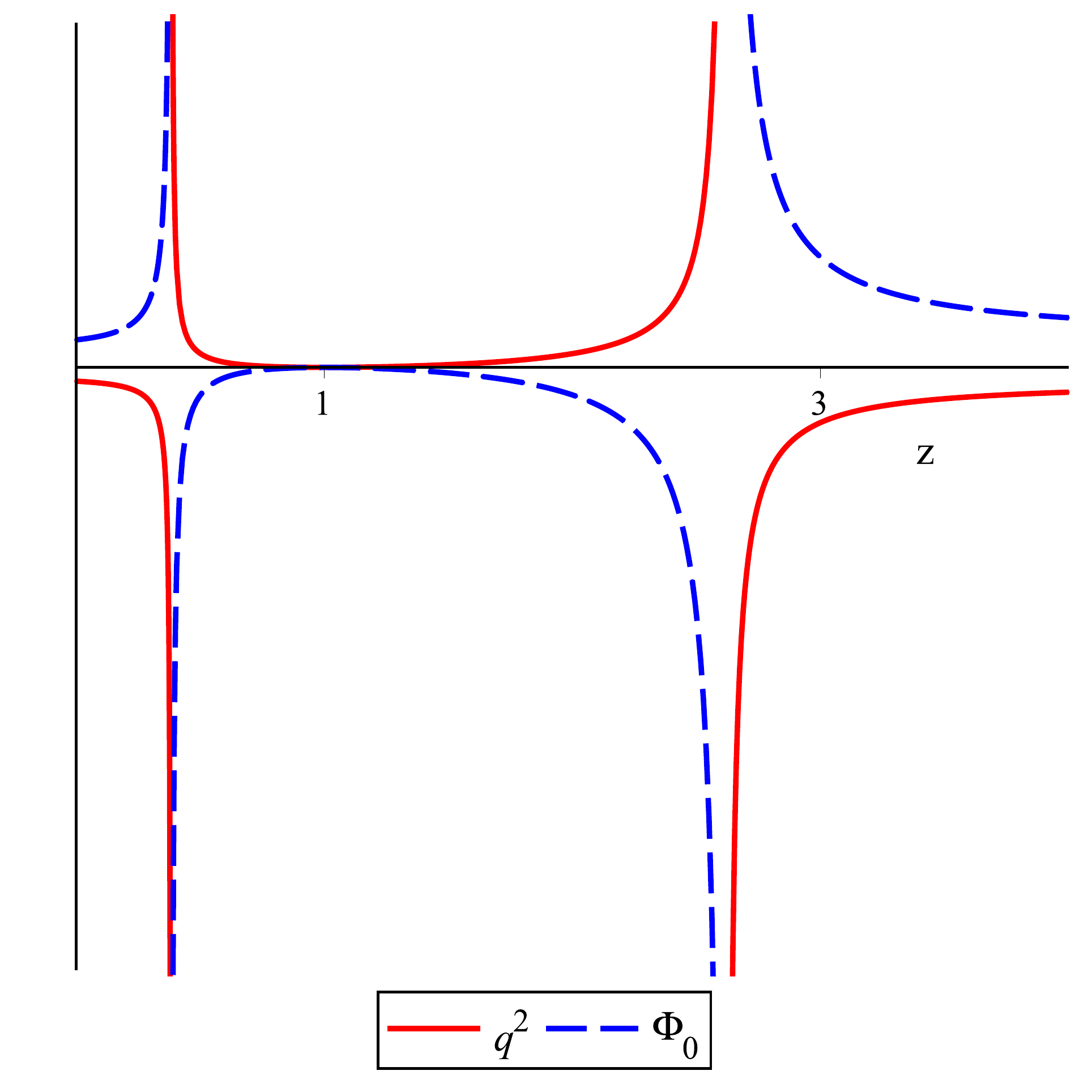}
      \caption{Representation of $q^2$ and $\Phi_0$ as functions of $z$, assuming $\kappa>0$.}
  \label{fig:q2vsphi0}
\end{figure}

The next case corresponds to setting the value $\chi=2$ in eq. (\ref{eq:ett-exx}). In this scenario,
the combination $E^t_{\, t}-E^{\varphi}_{\, \varphi}=0$
 with the appropriate value of $\Phi_0$ (shown in eq. (\ref{eq:phi0})), is proportional to
\begin{eqnarray}
0&=&\frac{2 (z-2) (z-3) (z-1)^2 (8 \xi-1)}{1-(z+5) \xi} \left(\frac{r_h}{r}\right)^4 \nonumber\\
&&-\Biggl\{
\frac{2(z-3) (z-1)^2 [(8 z+4) \xi-z-1] }{1-(z+5) \xi} \nonumber \\
&&+2\kappa l^4 (z^2-3 z+1)  q^2
\Biggr\} \left(\frac{r_h}{r}\right)^2.
\end{eqnarray}
From this expression, it is clear to see that the case $z=3$ recovers the uncharged solution found in \cite{AyonBeato:2009nh}.
One could also naively consider that there exists a solution for the case $\xi=1/8$, with $q^2 = -\frac{4(z-1)^2}{\kappa l^4(z^2-3z+1)}$ and {$\Phi_0 = \frac{8(z-1)^2}{\kappa(z^2-3z+1)}$}. However, we notice that there are no values of $z$ for which $q^2$ and $\Phi_0$ are simultaneously positive (see FIG. \ref{fig:q2vsphi0} for clarity). Therefore, this configuration does not represent a physical solution.

Nevertheless, setting the critical dynamical exponent $z=2$ in eq. (\ref{eq:ett-exx}) gives us a solution with an arbitrary non-minimal coupling $\xi$. The solution reads as follows
\begin{eqnarray}
ds^2&=&-\frac{r^{4}}{l^{4}} \left( 1- \frac{r_h^2}{r^2} \right) dt^2
+\frac{l^2}{r^2} \left( 1- \frac{r_h^2}{r^2} \right)^{-1} dr^2 \nonumber \\
&&+\frac{r^2}{l^2} d\varphi^{2}, \label{sol:chi2} \\
\Phi (r) &=& \sqrt{\frac{1}{\kappa(7\xi-1)}}\frac{r_h}{r}, \\
F_{rt}&=&\sqrt{\frac{(20\xi-3)}{\kappa(7\xi-1)l^4}}\, \, r_h,
\end{eqnarray}
supported by the self-interacting potential
\begin{eqnarray}
U(\Phi) &=& -\frac{\kappa\xi (7\xi-1)}{2l^2}\Phi^4+\frac{7\xi-1}{l^2}\Phi^2, \label{eq:potsol2}
\end{eqnarray}
with the following constants
\begin{eqnarray}
m^2 = \frac{1}{2l^2},  \qquad \lambda = -\frac{7}{2l^2}. \label{eq:constantsol2}
\end{eqnarray}
As before, to ensure that we have real values of
$$\Phi_0=\frac{1}{\kappa(7\xi-1)} \quad \mbox{and}
\quad q^2=\frac{(20\xi-3)}{\kappa(7\xi-1)l^4},$$
 we must consider the range of $\xi$ given by the TABLE \ref{tabla2}

\begin{table}[h!]
\begin{center}
\begin{tabular}{|c|c|}
\hline
$\kappa$ & Range of $\xi$ \\
\hline \hline
~$\kappa>0$~ & ~$\xi \in (3/20,+\infty)$~ \\  [1ex]
\hline \hline
~$\kappa<0$~ & ~$\emptyset$~ \\  [1ex]
\hline
\end{tabular}
\end{center}
\caption{\label{tabla2}Range of possibilities for non minimal coupling
parameter $\xi$ allowing black holes.}
\end{table}%

\begin{figure}[h!]
  \centering
    \includegraphics[scale=.42]{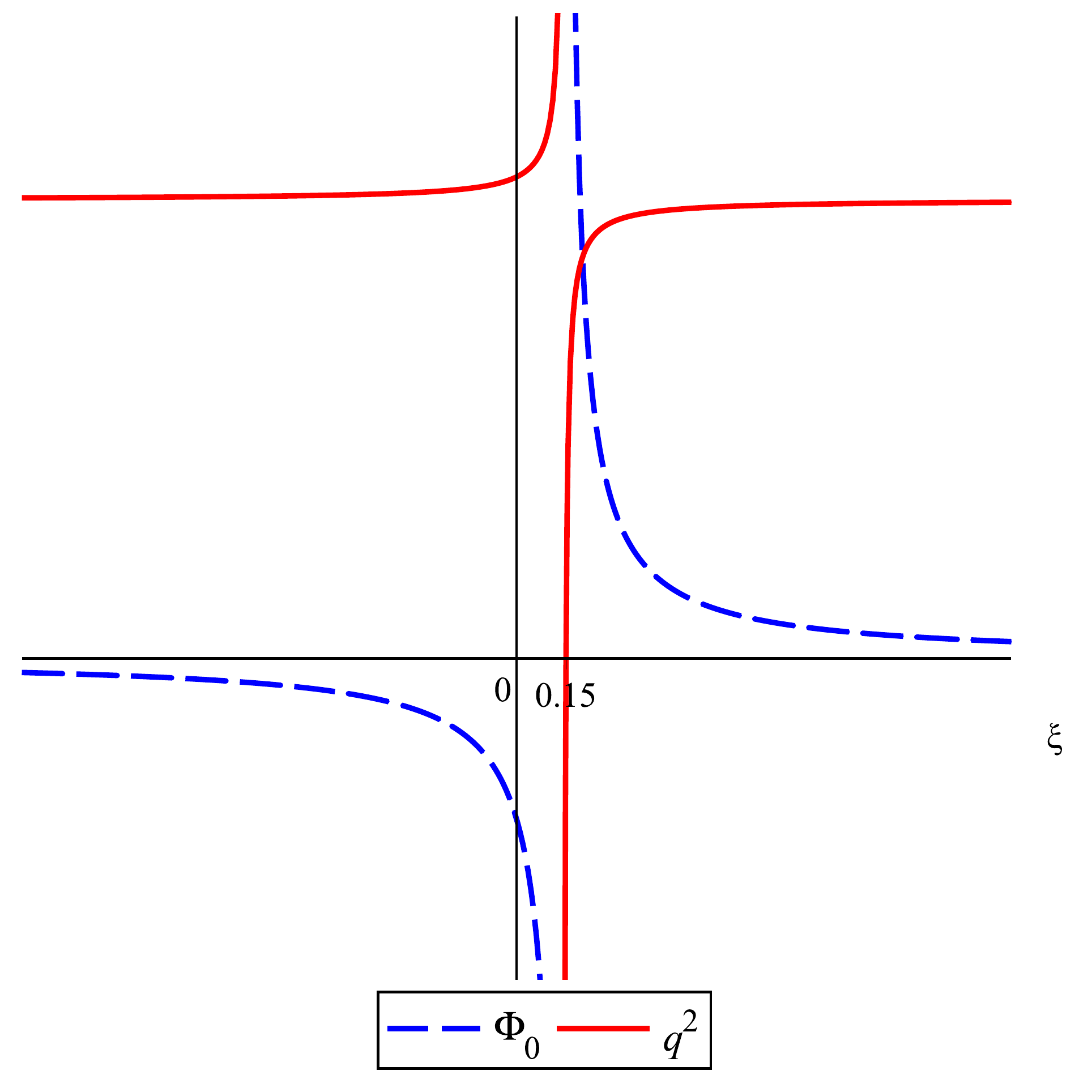}
      \caption{Representation of $q^2$ and $\Phi_0$ as functions of $\xi$ for the $\chi=2$ configuration, assuming $\kappa>0$.}
  \label{fig:q2vsphi0_chi2}
\end{figure}

Just for completeness, FIG. \ref{fig:q2vsphi0_chi2}  is provided as a visual aid on this matter. In fact, a quick inspection of the figure shows that for $\kappa>0$ the interval $\xi>3/20=0.15$ is the only one for which $\Phi_0$ and $q^2$ are simultaneously positive. Notice that taking $\kappa<0$ (equivalent to a reflection with respect to the horizontal axis) will not yield to any interval in which $\Phi_0$ and $q^2$ are simultaneously positive.

To summarize, we have found two families of charged black holes: one solution with an arbitrary dynamical exponent $z$ given by (\ref{sol:chi1})-(\ref{eq:potsol1}) whose coupling parameter $\xi$ is fixed as (\ref{xisol1}) and a solution with a fixed value of the dynamical exponent $z=2$ given by (\ref{sol:chi2})-(\ref{eq:potsol2}) that admits an arbitrary coupling parameter $\xi$ in the range $(3/20,+\infty)$.

\subsection{Corresponding solitons}
Considering that the other aim of this work is to check the consistency of a {charged Cardy-like expression, characterized by a magnetically charged soliton, we devote this subsection to achieve the task of building the form of the solitonic counterpart of the general solution
\begin{eqnarray}
ds^2=-\left(\frac{r}{l}\right)^{2z} \left[ 1- \left(\frac{r_h}{r}\right)^{\chi} \right] dt^2
+\frac{l^2}{r^2} \left[ 1- \left(\frac{r_h}{r}\right)^{\chi} \right]^{-1} dr^2 +\frac{r^2}{l^2} d\varphi^{2}\nonumber.
\end{eqnarray}
In order to achieve this we start by performing a double Wick rotation
\begin{equation}\label{eq:doublewick}
dt\mapsto id\varphi, \quad \quad d\varphi \mapsto idt.
\end{equation}
For the type of solutions that we are working with, we find that the solitons would be generally described by the following line element
\begin{eqnarray}\label{eq:soliton1}
ds^2=-\left( \frac{{r}}{l} \right)^2 dt^2+\frac{l^2}{r^2}\frac{d{r}^2}{f({r})}+\left(\frac{{r}}{l}\right)^{2z} f({r}) d\varphi^2,
\end{eqnarray}
with
\begin{eqnarray}
f({r})&=&1-\left[\left( \frac{2}{\chi}\right)^{1/z} \frac{l}{{r}} \right]^{\chi}, \nonumber\\
\Phi({r})&=&\sqrt{\Phi_0} \left[\left( \frac{2}{\chi}\right)^{1/z} \frac{l}{{r}} \right]^{\chi/2}, \nonumber \\
F_{{r} \varphi}&=&q l \left( \frac{2}{\chi}\right)^{1/z} \left( \frac{{r}}{l} \right)^{z-2},\label{eq:soliton2}
\end{eqnarray}
together with an adjustment of the horizon
$$r_h=l \left(\frac{2}{\chi}\right)^{1/z},$$
ensuring the correct identification with its Euclidean version. Here we note that, although its regular character is not manifest in this system of coordinates, it is a magnetically charged soliton. In the following section, as we analyse the thermodynamics of the black holes (\ref{sol:chi1}) and (\ref{sol:chi2}), we present the solitons corresponding to each case calculated using (\ref{eq:soliton1}).

\section{Thermodynamics of the solutions}\label{thermo}

From the previous section we note that the general expression for the charged black hole is determined {by} $\chi$, $\Phi_0$, $q$ and the dynamical exponent $z$. It is for this reason that in the following lines we derive  the generic versions of the thermodynamical quantities of the black hole of the form described in eq. (\ref{gensolLif_chi}),
these are the Wald entropy, the mass and electric charge.

\subsection{General thermodynamic quantities}

{As a first computation},  we calculate the Wald entropy $\mathcal{S}_{W}$ \cite{Wald:1993nt,Iyer:1994ys}
 \begin{eqnarray}
\label{wald} \mathcal{S}_{W}&\equiv&-2 \pi \oint_{\Sigma} d x \sqrt{|\gamma|} P^{a b c d} \, \varepsilon_{ab} \,  \varepsilon_{cd},
\end{eqnarray}
where $P^{abcd}=\delta {\mathcal{L}}/ \delta R_{a b c d}$ and the integral is to be taken at the spatial section $\Sigma$ of the event horizon, $|\gamma|$ denotes the determinant of the induced metric on $\Sigma$, $\varepsilon_{ab}$ is the binormal vector follows from the timelike Killing vector {$\partial_{t}=k^{\mu}\partial_{\mu}$}, which becomes null at the event horizon $r_h$ and reads
$$\varepsilon_{ab}=-\varepsilon_{ba}:=\frac{1}{\kappa}\,\nabla_{a} k_{b},$$
where the surface gravity $\kappa$ is given by
$$\kappa=\sqrt{-\frac{1}{2}\left(\nabla_{a} k_{b}\right)\left(\nabla^{a} k^{b}\right)}.$$
With all these ingredients, the Wald entropy (\ref{wald}) for ansatz (\ref{gensolLif_chi}) and the Lagrangian $\mathcal{L}$ defined in eq. (\ref{eq:Squad}) is given by
\begin{eqnarray}
\label{waldgensol} \mathcal{S}_{W}&=& - 2 \pi \, \Omega_{1} \,
\left(\frac{r_{h}}{\l}\right) \,
 \Big[P^{abcd} \, \varepsilon_{ab} \,  \varepsilon_{cd} \Big]_{r = r_{h}},
 \nonumber\\
&=&2\pi \Omega_{1} \left(\frac{r_h}{l}\right)\left[\frac{1}{\kappa}
-\Phi_0\xi-\frac{\chi(\chi+2-3z)}{4 \kappa  m^2 l^2}\right],
\end{eqnarray}
where $\Omega_{1}= 2 \pi l$, while the {Hawking} temperature for these configurations reads
\begin{equation}
T\equiv\frac{\kappa}{4 \pi}=\frac{r_h^{z+1}}{4\pi l^{z+1}}f'(r_h)=\frac{\chi}{4\pi l}\left(\frac{r_h}{l}\right)^z.
\end{equation}
The electric charge can be calculated by using
\begin{equation}
\mathcal{Q}_e\equiv\int_{\Sigma} d x \sqrt{|\gamma|} n^{\mu} u ^{\nu} F_{\mu \nu}=\int d\Omega_1 \left(\frac{r}{l}\right)^{2-z} F_{rt}=q r_h \Omega_1,
\end{equation}
where $|\gamma|$ is the induced metric on $\Sigma$, $n^{\mu}$ and $u^{\nu}$ are unit spacelike and timelike normals to $\Sigma$
$$u^{\nu}=\frac{l^{z}}{r^{z} \sqrt{f}}dt,\quad n^{\mu}=\frac{r \sqrt{f}}{l} dr,$$
and the electric potential is defined as
\begin{equation}
\Phi_e\equiv-A_{t}(r_h)=-\frac{q l^2}{z-1}\left(\frac{r_h}{l}\right)^{z}.
\end{equation}
On the other hand, the double Wick rotation (\ref{eq:doublewick}) and the correct identification of its Euclidean version yield the corresponding magnetic charge and potential for the soliton given by
\begin{eqnarray}
\mathcal{Q}_m&=&iql\left(  \frac{2}{\chi} \right)^{1/z}\Omega_1,\\
\Phi_m&=&-\frac{iq l^2}{z-1}\left(\frac{2}{\chi}\right),
\end{eqnarray}
where the values of the magnetic charge and potential are both purely imaginary. However, the product of the charge times the potential is always real.
Next, we derive a general expression for the mass, using the approach described in \cite{Kim:2013zha}, \cite{Gim:2014nba}. This method consists of a quasilocal generalization of the Abbott-Deser-Tekin (ADT) formalism \cite{Abbott:1981ff,Deser:2002rt,Deser:2002jk}, which ensures its reliability even when considering a Lifshitz asymptote. In order to proceed, we must first calculate two objects. The first is the Noether potential
\begin{eqnarray}
K^{\mu\nu}=\sqrt{-g}(2P^{\mu\nu\rho\sigma}\nabla_{\rho}k_{\sigma} - 4k_{\sigma}\nabla_{\rho}P^{\mu\nu\rho\sigma} +F_{\mu\nu}k^{\sigma}A_{\sigma}),
\end{eqnarray}
where, as before, $\mathcal{L}$ is the Lagrangian defined in eq. (\ref{eq:Squad}). The second object is the surface term given by
\begin{eqnarray}
\Theta^{\mu}&=&2\sqrt{-g}\Biggl(P^{\mu\alpha\beta\gamma}\nabla_{\gamma}\delta g_{\alpha\beta}-\delta g_{\alpha\beta} \nabla_{\gamma}P^{\mu\alpha\beta\gamma}+\frac{1}{2}\frac{\partial \mathcal{L}}{\partial(\partial_{\mu}\Phi)}\delta\phi\nonumber\\
&-&\frac{1}{2}F^{\mu\nu}\delta A_{\nu}\Biggr).
\end{eqnarray}
With these ingredients, a parameter $0\leq s \leq 1$ is introduced, which will allow the interpolation between the black hole solution with $s=1$ and the asymptotic one at $s=0$. As a result, the quasilocal conserved charge is given by
\begin{eqnarray}
\mathcal{M}(k)=\int_B dx_{\mu\nu}\left(\Delta K^{\mu \nu}(k)-2k^{[\mu} \int^{1}_0 ds \Theta^{\nu]}(k|s)\right),
\end{eqnarray}
where $\Delta K^{\mu\nu}(k)\equiv K^{\mu\nu}_{s=1}(k)-K^{\mu\nu}_{s=0}(k)$ is the difference of the Noether potential between the interpolated solutions.
In short, applying this technique to our particular case, the general equation to calculate the mass for black holes of the form eq. (\ref{gensolLif_chi}) is given by
\begin{eqnarray}
\label{massgensol}
\mathcal{M}(k)&=&
\Omega_1 \Bigg\{
-\frac{ q^2r^{z-1} }{2(z-1) l^{z-4}}\left(\frac{r_h}{l}\right)^{2}\nonumber \\
&+& \left[\frac{\Xi_1}{2 \kappa}-\frac{ \Phi_{0} l^{2}}{4}\left( \left( 1-2\,\chi-2\,z
\right) \xi+\frac{\chi}{2} \right)
\right]\,\left(\frac{r_h}{l}\right)^{2 \chi}\, \frac{r^{1+z-2
\chi}}{l^{4+z-2 \chi}}\,
\nonumber\\
&+& \left[\frac{\Xi_2}{2 \kappa} -\Phi_{0} l^{2}\left(
\xi\,z+\xi\,\chi-\frac{\chi}{4} \right)
\right]\,\left(\frac{r_h}{l}\right)^{\chi}\,\frac{r^{1+z-
\chi}}{l^{4+z-\chi}}\Bigg\}
,
\end{eqnarray}
where
\begin{eqnarray*}
\Xi_1&=&\frac{1}{8 m^{2}} [2\chi^3-(1+4z)\chi^2-(10+2z^2-15z)\chi \\
&&+2(2z-1)(z^2-3z+1)],\\
\Xi_2&=&\frac{1}{2 m^{2}} [-\chi^3+2 z \chi^2+(4+z^2-6 z)\chi \\
&&-(2z-1)(z^2-3z+1)+2l^2m^2].
\end{eqnarray*}

Similarly to the case of the black holes, the general expression for the mass of the corresponding soliton is given by
{
\begin{eqnarray}
 \label{massgensoliton}
\mathcal{M}_{\mathrm{sol}}(k)&=&
\Omega_1 \Biggl\{
\frac{  q^2{r}^{z-1} }{2(z-1) l^{z-4}}\left(\frac{2}{\chi}\right)^{2/z} \nonumber\\
&+& \left[ \frac{\Psi_{1}}{ 2
\kappa}-{\Phi}_0 l^{2}\left(\frac{{\chi}-2\xi\, \left(
4\chi-2\,z+3 \right)}{8}
 \right)
 \right]\,\left(\frac{2}{\chi}\right)^{2\chi/z}
 \frac{{r}^{1+z-2\chi}}{{l^{4+z-2\chi}}}
 \nonumber \\
&+&\left[\frac{\Psi_{2}}{ 2 \kappa}- {\Phi}_{0} l^{2} \left(
\xi(1+\chi)-\frac{\chi}{4} \right)
\right]\,\left(\frac{2}{\chi}\right)^{\chi/z}
\frac{{r}^{1+z-\chi}}{{l^{4+z-\chi}}} \Biggr\},
\end{eqnarray}
}
where
\begin{eqnarray*}
\Psi_1&=&\frac{1}{8 m^{2}} [-4\chi^3+(10 z-1)\chi^2-(2z^2+17z-18)\chi\\
&&-2(2z-3)(z^2-3z+1)],\\
\Psi_2&=&\frac{1}{2 m^{2}} [\chi^3+(1-3z)\chi^2+\big((z+3)(z-1)-2m^2l^2\big)\chi\\
&&+2(2z-1)
l^2m^2+(2z-3)(z^2-3z+1)].
\end{eqnarray*}
 Notice that the general expressions (\ref{massgensol}) and (\ref{massgensoliton}) from this quasilocal formulation \cite{Kim:2013zha,Gim:2014nba} are obtained along a one-parameter family of configurations and for actual solutions must not depend on the radial coordinates $r$. Interestingly, this gives an indication of constraints on the constants $z,q,\chi,\Phi_0, \Xi_i$ and $\Psi_i$, being consistent with the two families of charged Lifshitz black holes found previously in Section \ref{section:eom}.\\

In the next subsections, we analyze, in depth, the thermodynamics of the two new families of charged Lifshitz black holes solutions where, additionally, we will check that their Wald entropy (\ref{waldgensol}) is correctly reproduced by means of {a charged Cardy-like formula proposed in \cite{Bravo-Gaete:2015iwa} and given by
\begin{eqnarray}\label{cardycharged}
\mathcal{S}_{C}&=&2 \pi l (z+1)\left( |-\mathcal{M}_{\mathrm{sol}}z^{-1}+\alpha \Phi_m \mathcal{Q}_m|^z   |\mathcal{M}-\alpha \Phi_e \mathcal{Q}_e| \right)^{\frac{1}{z+1}},
\end{eqnarray}
where the constant $\alpha$ depends on the electromagnetic Lagrangian considered and it is related with the three dimensional Smarr formula \cite{Smarr:1972kt}
\begin{equation}\label{Smarrcharged}
\mathcal{M}=\left(\frac{1}{z+1}\right)T\mathcal{S}+\alpha \mathcal{Q}_e \Phi_e,
\end{equation}
which in our case takes the value
\begin{equation}\label{alpha}
\alpha=\frac{1}{z+1}.
\end{equation}
Note that {\em{a priori}} we used the notation ${\cal{S}}_{C}$ in order to differentiate with respect to the Wald entropy ${\cal{S}}_{W}$. Nevertheless, as we will see in the examples of the following subsections, both thermodynamic quantities are the same, this is ${\cal{S}}_{C}={\cal{S}}_{W}$.} Moreover, with all the calculations provided, we will verify that the first law of thermodynamics
\begin{equation}\label{firstlaw}
d\mathcal{M}=Td\mathcal{S}+\mathcal{Q}_e d \Phi_e,
\end{equation}
holds.

With all the above, we are now ready to analyze, explicitly, the thermodynamics of the solutions that we obtained in section \ref{section:eom} in the following subsections.

\subsection{ Thermodynamics of the solution with arbitrary dynamical exponent $z$. \label{1stclass}}

 For the first family of charged Lifshitz black holes (\ref{sol:chi1})-(\ref{eq:potsol1}) characterized by the coupling constants (\ref{eq:m2})-(\ref{eq:lambda}) together with the nonminimal coupling parameter $\xi$ (\ref{xisol1}) and using eq.({\ref{waldgensol}}) we find that this configuration has a non-vanishing Wald entropy given by
\begin{eqnarray}\label{Waldentropysoln1}
\mathcal{S}_{W}=-{\frac {2 \pi z \left( 2\,z-3 \right) \Omega_1}{\kappa
 \left( z-1 \right)  \left( {z}^{2}-3\,z+1 \right)}} \left(\frac{r_h}{l}\right),
\end{eqnarray}
which is positive when $z\in (3/2,2)$ and $\kappa>0$,
while its temperature is
\begin{eqnarray}
T&=&\frac{1}{4 \pi l} \left(\frac{r_h}{l}\right)^{z},
\end{eqnarray}
and the electrical quantities read
\begin{eqnarray}
\Phi_{e}&=&-\frac{1}{2 (z-1)} \sqrt{-{\frac {2 z \left( 2\,z-3 \right)}
{ \left( {z}^{2}-3\,z+1 \right) \kappa}}
}\left(\frac{r_h}{l}\right)^{z} , \\
\mathcal{Q}_{e}&=&\frac{\Omega_1}{2\,l} \sqrt{-{\frac {2 z \left( 2\,z-3 \right)}
{ \left( {z}^{2}-3\,z+1 \right) \kappa}}
}\left(\frac{r_h}{l}\right).
\end{eqnarray}
Using eq. (\ref{massgensol}), we find that the mass $\mathcal{M}$ vanishes. With all the above, we check that the first law of black hole thermodynamics (\ref{firstlaw}) as well as the Smarr formula (\ref{Smarrcharged}) {hold}, and this solution can be interpreted as an extremal charged Lifshitz black hole in the sense that it has zero mass while having a non-vanishing entropy and electric charge. {We note that this feature is not new, in fact, charged configurations with this characteristic have been found previously in \cite{Pang:2009pd} with an Einstein-Maxwell-Proca system and in higher dimensions with a Einstein-Maxwell toy model together with the most general quadratic corrections of the gravity \cite{Bravo-Gaete:2015xea}.} Additionally, one could also calculate the corresponding soliton as described before and compute its conserved charges. Indeed, we find that the corresponding soliton is given by
\begin{eqnarray}
ds^2&=&-\left(\frac{{r}}{l}\right)^{2}  dt^2
+\frac{l^2}{{r}^2} \left( 1-2^{1/z} \frac{l}{{r}} \right)^{-1} d{r}^2 \nonumber\\
&&+\left(\frac{{r}}{l}\right)^{2z} \left( 1-2^{1/z} \frac{l}{{r}} \right) d\varphi^2,
\end{eqnarray}
with
\begin{eqnarray*}
\Phi({r})&=&\sqrt{{ \frac{ 8(z^2-z+1)(z-2)}{(z-1)(z^2-3z+1)\kappa}}}\, \,  2^{\frac{1}{2z}}  \sqrt{\frac{l}{{r}}},\nonumber\\
F_{{r}\varphi}&=&\sqrt{-\frac{z(2z-3)}{2\kappa(z^2-3z+1)l^2}}\, \,  2^{1/z} \frac{{r}^{z-2}}{l^{z-2}}.
\end{eqnarray*}
Using eq. (\ref{massgensoliton}), we find that the mass of the soliton vanishes, as expected, and the rest of the corresponding quantities of interest for this magnetically charged soliton are given by
\begin{eqnarray}
\Phi_{m}&=&=-\left(\frac{i}{z-1}\right) \sqrt{-{\frac {2 z \left( 2\,z-3 \right)}
{ \left( {z}^{2}-3\,z+1 \right) \kappa}}} ,\nonumber\\
\mathcal{Q}_{m}&=&\frac{ 2^{\frac{1-z}{z}}\,i \,\Omega_1}{l}\, \sqrt{-{\frac {2 z \left( 2\,z-3 \right)}
{ \left( {z}^{2}-3\,z+1 \right) \kappa}}
},
\end{eqnarray}
where it is easy to see that the formula (\ref{cardycharged}) with the constant $\alpha$ (\ref{alpha})  correctly fits the Wald entropy $\mathcal{S}_{W}$ given previously in (\ref{Waldentropysoln1}).

\subsection{Solution with a fixed value of the dynamical exponent $z=2$.}
 Regarding the second family of solutions (\ref{sol:chi2})-(\ref{eq:potsol2}) with the constants fixed as (\ref{eq:constantsol2}), using eq. (\ref{waldgensol}) we find that the Wald entropy of this black hole is given by
\begin{eqnarray}\label{Waldentropysoln2}
\mathcal{S}_{W}=\frac{2 \pi (20\xi-3) \Omega_1}{\kappa (7\xi-1)}\left(\frac{r_h}{l}\right),
\end{eqnarray}
which in this case is always a positive quantity, while its temperature is
\begin{eqnarray}
T=\frac{1}{2 \pi l} \left(\frac{r_h}{l}\right)^{2}.
\end{eqnarray}
At the same time, its electrical quantities are given by
\begin{eqnarray}
\Phi_{e}&=&-\sqrt{\frac{20\xi-3}{\kappa (7\xi-1)}}\left(\frac{r_h}{l}\right)^{2} ,\nonumber\\
\mathcal{Q}_{e}&=&\frac{\Omega_1}{l} \sqrt{\frac{20\xi-3}{\kappa (7\xi-1)}}\left(\frac{r_h}{l}\right),
\end{eqnarray}
and, using eq. (\ref{massgensol}) we find that $\mathcal{M}=0$ and this solution also corresponds to an extremal black hole in the same sense than in the previous case. It is simple to verify that the first law (\ref{firstlaw}) and the Smarr formula (\ref{Smarrcharged}) are satisfied.

Furthermore, in this case, we can repeat the double Wick rotation process described earlier in this section to find the corresponding soliton which, for this solution, is given by
\begin{eqnarray}
ds^2&=&-\left(\frac{{r}}{l}\right)^{2}  dt^2
+\frac{l^2}{{r}^2} \left[ 1- \frac{l^2}{{r}^2} \right]^{-1} d{r}^2 \nonumber\\
&&+\left(\frac{{r}}{l}\right)^{4} \left[ 1- \frac{l^2}{{r}^2} \right] d\varphi^2,
\end{eqnarray}
with
\begin{eqnarray}
\Phi({r})&=&\sqrt{{\frac{1}{\kappa(7\xi-1)}}}\left( \frac{l}{{r}} \right),\\
F_{{r}\varphi}&=& \sqrt{\frac{(20\xi-3)}{\kappa(7\xi-1)l^2}}\, \,  .
\end{eqnarray}
In this case, using eq. (\ref{massgensoliton}), we find that the mass of the soliton is null, as expected, and the magnetically charged soliton's quantities read
\begin{eqnarray}
\Phi_{m}&=&-i\sqrt{\frac{20\xi-3}{\kappa (7\xi-1)}} ,\nonumber\\
\mathcal{Q}_{m}&=&\frac{i \Omega_1}{l} \sqrt{\frac{20\xi-3}{\kappa (7\xi-1)}}.
\end{eqnarray}
{With this information, it is straightforward to check} that the formula (\ref{cardycharged}) with $z=2$ and $\alpha=1/3$ coincides with the Wald entropy (\ref{Waldentropysoln2}), that is, $\mathcal{S}_{W}=\mathcal{S}_{C}$.

\section{Discussion and conclusions}\label{section:conclusions}

The aim of this work was, first, to find charged asymptotically Lifshitz black holes using a linear Maxwell field in New Massive Gravity with a non-minimally coupled scalar field and to analyze their {thermodynamic quantities. Indeed, for this theory}  we were able to find two solutions, where {their
metric functions depend only on an integration constant and one} of them does not require the dynamical exponent $z$ to be fixed, while the other one exists only for $z=2$. Our second motivation was to provide new examples for the use of the quasilocal formalism proposed in \cite{Kim:2013zha}, \cite{Gim:2014nba} when considering two types of matter contributions (the linear Maxwell field and the scalar field). Regarding this, we were able to {find new examples of extremal charged configurations} in the sense that, while they both have vanishing mass, they possess non-zero entropy and non-zero electric charge. Moreover, with all the computations, we verify that the first law (\ref{firstlaw}) is held and that our solutions comply with the Smarr formula (\ref{Smarrcharged}), which is in correspondence with a more general expression in $D$ dimensions
\begin{equation*}
\mathcal{M}=\frac{D-2}{z+D-2}\left(T\mathcal{S}+\mathcal{Q}_e \Phi_e\right),
\end{equation*}
proposed in \cite{Dehghani:2013mba}.

It is interesting to note that, { given these extremal thermodynamic behaviour, where the mass of the charged Lifshitz black hole as well as its solitonic configuration vanishes, the entropy of these solutions can be computed by means of a charged Cardy-like expression (\ref{cardycharged}) proposed in \cite{Bravo-Gaete:2015iwa} with $\alpha$ given by (\ref{alpha}), where the ground state is given by a magnetically charged soliton obtained by a double Wick rotation, checking that this formula} fits perfectly with the Wald entropy $\mathcal{S}_{W}$ previously computed in both cases. Also notice that although the magnetic charge and potential of the soliton are both purely imaginary, the product $\Phi_m \mathcal{Q}_m $ that appears in the Cardy-like formula above is real, which allows an appropriate physical interpretation.

Some natural extensions of this work would be to extend this approach to higher dimensions \cite{Bra-Ju} or, complementarily, to consider a more general asymptotic metric in which the scaling transformation does not act as an isometry but rather like a conformal transformation. This type of metrics are commonly referred to as hyperscaling violation metrics and are, in general, described by \cite{Charmousis:2010zz}
\begin{equation*}
ds^2_{H}=\left( \frac{l}{r} \right)^{\frac{2\theta}{D-2}}\left( -\frac{r^{2z}}{l^{2z}}dt^2+\frac{l^2}{r^2}dr^2+\frac{r^2}{l^2} d\vec{x}^{2}  \right),
\end{equation*}
where $\theta$ is known as the hyperscaling violation exponent.
Another interesting open problem would be to find non-extremal charged solutions that can be supported by a Maxwell field, perhaps considering more diverse matter contributions associated to scalar fields such as the Horndeski action which is the most general tensor scalar action that yields to second order field equations in four dimensions \cite{Horndeski:1974wa} or, conversely, exploring gravitational actions complemented with non-minimally coupled scalar fields plus a more general component for the electrodynamics contribution.

\section*{Acknowledgments}
We thank Eloy Ay\'on-Beato, Mokhtar Hassa\"ine, Gerardo Vel\'azquez and Benito Ju\'arez for useful discussions. MB is supported by grant Conicyt/ Programa Fondecyt
de Iniciaci\'on en Investigaci\'on No. 11170037. MMJA would like to thank Universidad Cat\'olica del Maule for their hospitality. MB would like to thank Arkansas State University Campus Quer\'etaro for their hospitality.



\end{document}